\documentclass[amsmath,amssymb,aps,twocolumn]{revtex4-2}

\usepackage{slashed}
\usepackage{amsmath}
\usepackage{mathrsfs}
\usepackage{graphicx}
\usepackage{dcolumn}
\usepackage{bm}
\usepackage{hyperref}
\usepackage{subfigure}
\usepackage{caption}
\usepackage{amsfonts,amssymb}
\usepackage[utf8]{inputenc}
\usepackage{pstricks}
\usepackage{color}

\newcommand{\be}{\begin{equation}}
\newcommand{\ee}{\end{equation}}
\newcommand{\ben}{\begin{eqnarray}}
\newcommand{\een}{\end{eqnarray}}

\begin{document}


\title{{ An Effective Field Theory Treatment of the Production and Annihilation of Magnetic Monopoles and their Relic Abundance}}

\author{Luciano M. Abreu}
 \email{luciano.abreu@ufba.br}
\affiliation{Instituto de Física, Universidade Federal da Bahia, Campus Ondina, 40170-115 Salvador, Bahia, Brazil}

\author{Pedro C. S. Brandão}
\email{pedro.brandao@ufba.br}
\affiliation{Instituto de Física, Universidade Federal da Bahia, Campus Ondina, 40170-115 Salvador, Bahia, Brazil}

\author{Marc de Montigny}
\email{mdemonti@ualberta.ca}
\affiliation{Faculté Saint-Jean, University of Alberta, Edmonton, Alberta, T6C 4G9, Canada}

\author{Pierre-Philippe A. Ouimet}
\email{Pierre-Philippe.Ouimet@uregina.ca}
\affiliation{Department of Physics, University of Regina, Regina, Saskatchewan, S4S 0A2, Canada}

\begin{abstract}

We revisit the thermal production red and annihilation of magnetic monopoles and their relic abundance in order to gain a deeper physical interpretation on the monopole phenomenology predicted from the  Baines {\it et al}'s effective field theory, recently proposed in the description of monopole pair production via Drell-Yan and photon fusion processes. In this sense,  we red use of the vacuum cross sections for the Drell-Yan reactions derived within the mentioned framework to evaluate the cross section averaged over the thermal distribution associated to other  particles that constitute the hot medium where the monopoles propagate. In the considered range of monopole mass with spin-zero and spin-half, our findings suggest that the thermally averaged cross sections for the pair production are highly suppressed, while at higher temperatures those for the annihilation of lighter pairs  reach larger magnitudes. Besides, we observe that smaller temperature leads to a rate of annihilation for scalar monopoles smaller than the one for fermionic monopoles, which might be interpreted as a theoretical evidence of a more pronounced stability for spin-$0$ and heavier monopoles. Then we input these thermally averaged cross sections into the kinetic equation that describes the evolution of the monopole abundance via an extension of a freeze-out theory. Our  results infer that heavier monopoles achieve the equilibrium at earlier stages of the expansion, and consequently at higher temperatures. In  addition, larger monopole masses produce higher values of the relic abundance. Besides, the results indicate that the abundance does not behave differently for spinless and spin-half relic monopoles. 
\end{abstract}

\maketitle

\section{\label{sec:level1}Introduction}

The existence of magnetic monopoles, first introduced by Dirac, offers an elegant explanation for the quantization of electric charge as well as being a feature of several beyond-the-Standard-Model theories. While these objects have been investigated theoretically in the past, no experimental evidence of their existence has yet been found~\cite{Dirac, tHooft, Polyakov, Polchinski,Mavromatos:2020gwk}.  

With Run 3 of CERN's LHC, expected later this year, about to push the energy frontier further, the possibility of direct detection of a magnetic monopole is once more in the cards. In addition to searches at ATLAS \cite{atlas} the dedicated MoEDAL experiment will be taking data during this run. This experiment, located at the Interaction Point 8 on the LHC ring at CERN ~\cite{Moedal1} has published some of the most recent bounds on monopole masses and charges. Other monopole searches include Refs. \cite{icecube,icecube2,icecube3,kamiokande,anita,nova}.

Recently, the analysis of the MoEDAL trapping detector provided mass limits in the range 1500-3750 GeV for magnetic charges up to 5$g_D$ for monopoles of spin 0, 1/2 and 1 ~\cite{Moedal2}. A more recent search, for magnetic monopole production by the Schwinger mechanism in Pb-Pb heavy ion collisions at the LHC, excluded monopoles up to 75 GeV/$c^2$ to 70 GeV/$c^2$  for magnetic charges from 1$g_D$ to 3$g_D$, respectively, at a  95\% confidence level ~\cite{Moedal3}.  That work exploited the idea that the production of monopole-antimonopole pairs are most probable within a strong magnetic fields created in heavy-ion collisions such as the high-energy Pb-Pb collision at the LHC~\cite{Gould}. Another active topic of research is the possibility of detection of bound states of monopole pairs, or monopolia ~\cite{Abreu:2020qqy,Mavromatos:2020gwk,Epele,Vento1,Barrie,Reis,Vento2}.
	 
The new theoretical interest in monopoles, motivated by active experimental searches at higher energies, has stimulated new theoretical approaches to monopole physics. One significant theoretical challenge is the potential large value of the magnetic charge and, as a consequence, the non-perturbative nature of the corresponding field theory. In order to overcome this challenge, Baines {\it et al} introduced an effective monopole theory with a velocity-dependent magnetic charge ~\cite{Baines2018}, in  which slower monopoles would see a suppression of their interactions, thus leading to a perturbative theory for these monopoles. This effective theory is a U(1) gauge field theory which is motivated by using electric-magnetic dualisation. Baines {\it et al} considered the production of point-like monopoles of spin 0, 1/2 and 1 via the photon-fusion and Drell-Yan processes  ~\cite{Baines2018}. 
	
The absence of observation of magnetic monopoles at the LHC up to  this date, coupled with the non-observation of any relic monopoles argues that the relic monopole density is small. As a deeper investigation of Baines {\it et al}'s effective theory, we will investigate the  behavior of this description in the early universe. Along those lines, we examine the behavior of spinless and spin-half magnetic monopole pairs in a high-energy environment via the Drell-Yan process. 
		
An early discussion of monopole pair production from thermalized quark-gluon matter with the fireball model in heavy-ion collisions is in Refs. ~\cite{Roberts, Dobbins}. More recently,  the authors of Ref. ~\cite{Rajantie} investigated lower bounds on the mass of magnetic monopoles in heavy-ions via the Schwinger process, relying on the rich production of monopoles afforded by strong magnetic fields and high temperatures.   In  Ref. ~\cite{Gould}, they also examined the Schwinger production of monopoles in peripheral heavy-ion collisions with the use of the wordline instanton method to all orders in the magnetic charge.
	 
Since we are interested in the monopole freeze-out using the effective theory of Baines {\it et al}, in order to compute the relevant cross sections, we exploited results in Ref. ~\cite{Baines2018}.  In this paper, we obtain the thermal-averaged cross sections for monopole absorption, in analogy with recent work on the open flavour tetraquark state X(2900) in a hot hadronic medium Ref. ~\cite{Abreu2021,Abreu2022}.  We obtained the thermally averaged cross section $<\sigma v_{rel}>$ for the absorption of monopoles with an exact formula obtained by Cannoni and which is valid with an effective field theory where the masses of the annihilation products (here, the quarks) are much smaller than the monopole masses ~\cite{Koch,Cannoni2016,Cannoni2014}. 

Equipped with the thermally averaged cross sections, we then turn to the monopole abundance by exploiting an approach utilized to describe the time evolution of the number density of weakly interacting massive particles (WIMP) candidates for dark matter following Refs. ~\cite{Jungman,Zeldovich,Weinberg1977}. Therein, as in our present discussion of monopoles, when the Universe cools, the equilibrium abundance decreases until the annihilation reaction rate falls behind the Hubble expansion rate, where the interactions maintaining thermal equilibrium freeze out.  We apply the formalism of Ref. ~\cite{Cannoni2015}  for the time evolution of the abundance of magnetic monopoles.

In Sec. \ref{s2}, we briefly describe our effective Lagrangians for spinless and spin-half monopoles, for the production and annihilation of monopole-antimonopole pairs via the Drell-Yan processes, as well as the related cross sections. Then we compute the thermally-averaged cross sections for the production and annihilation of monopole-antimonopole pairs in Sec. \ref{s3}. This allows us to investigate the relic abundance of monopoles, in terms of the monopole mass and the temperature, in Sec. \ref{s4}. Lastly, we present concluding remarks in Sec. \ref{s5}.

\section{The effective formalism\label{s2}}

In order to investigate the  production and annihilation mechanisms of monopoles required to assess the evolution of their abundance, we consider their electromagnetic interactions with ordinary photons. However, the fundamental theory of monopoles, if any, is not firmly established. In that regard, as in other studies, here we adopt the following ansatz: the use of effective field-theoretical models based on electric-magnetic duality. More concretely, we employ an effective $U(1)$ gauge field theory introduced in Ref.~\cite{Baines2018} which describes the interaction of monopole fields with photons, by replacing the electric charge $q_e$ by the magnetic charge $g_D$. Accordingly, $g_D$ is quantized as required by Dirac’s quantisation rule, and provides the magnitude of the interaction of magnetically-charged fields with ordinary photons. 

With these guiding principles, and keeping the monopole's spin as a free parameter, we recall from Ref.~\cite{Baines2018} the effective Lagrangians for spinless monopoles $\phi$ and spin-half monopoles $\psi$, with the photon gauge field:
\begin{eqnarray}
{\mathcal{L}^{(S=0)}} & = & -\frac{1}{4}F^{\mu\nu}F_{\mu\nu} + (D^{\mu}\phi)^{\dagger}D_{\mu}\phi - M^2\phi^{\dagger}\phi , \nonumber \\
{\mathcal{L}^{(S=1/2)}} & = & -\frac{1}{4}F^{\mu\nu}F_{\mu\nu} + \overline{\psi}(i\slashed{D} - M)\psi. 
\label{Lagr}
\end{eqnarray}
$M$ is the monopole mass; $F_{\mu\nu}\equiv \partial_\mu A_\nu -\partial_\nu A_\mu$ and
 $D_\mu\equiv \partial_\mu-i g_D A_\mu$ are the field strength tensor and covariant derivative associated to the $U(1)$ gauge field $ A_\mu$ (i.e. the photon field). 
 
As a first attempt, we devote our attention to the lowest-order diagrams contributing to the monopole pair production and suppression $q\bar{q}\rightarrow M\bar{M}$, i.e. Drell-Yan processes seen as monopole-antimonopole pair production from a quark-antiquark annihilation (and their respective inverse ones). These reactions have already been studied in Ref.~\cite{Baines2018}, and we do not reproduce the details here, for the sake of conciseness. Notwithstanding, for completeness, we introduce some relevant quantities and information necessary to the subsequent sections. We remark that although this calculation is done with the quark-antiquark annihilation, it can be generalized for other fermionic fields of the standard model (like the electrons and positrons), by imposing the appropriate coupling describing the interaction of the fermions with the photon and adding a correction factor associated to the relevant internal degrees of freedom (i.e. for quarks they are related to the color and flavor states). We think that these should not display major qualitative changes compared to our findings reported below. 

The spin-averaged cross section in the center of mass (CM) frame for the processes mentioned above is given by
\begin{equation}
\sigma_{ab \rightarrow cd} ^{(S)} (s) = \frac{1}{64 \pi^{2} s} \frac{ |\vec{p}_{cd}| }{ |\vec{p}_{ab}|} \int d \Omega \overline{\sum_{S}} | \mathcal{M}_{ab \rightarrow cd}^{(S)} (s, \theta) |^{2} , \label{EqCrSec}
\end{equation}
where $\sqrt{s}$ is the CM energy;  $|\vec{p}_{ab}|$ and $|\vec{p}_{cd}|$
stand for the three-momenta of initial and final particles in the CM frame,
respectively; $ \mathcal{M}_{ab \rightarrow cd} (s, \theta)$ denotes the sum of the transition amplitudes of all processes contributing to the interaction; and 
the symbol ${\overline{\sum}}_{S}$ represents the sum over the spins (or polarizations and colors, as needed) of the particles in the initial and final state, weighted by the degeneracy factors $g_{1i}$ and $g_{2i}$  of the two particles forming the initial state.

Then, based on the effective Lagrangians introduced in Eq.~(\ref{Lagr}), the amplitudes for the monopole pair production via Drell-Yan processes can then be calculated, and after some manipulations we get the total cross sections taking the limit of negligible quark (or other fermions) masses when compared to heavy monopoles~\cite{Baines2018}:
\begin{eqnarray}
{\sigma^{(S=0)}_{q\bar{q}\rightarrow M\bar{M}}(s)} &  = & \frac{5\pi\alpha_g\alpha_e}{27s}\beta^3 \nonumber \\
{\sigma^{(S=1/2)}_{q\bar{q}\rightarrow M\bar{M}}(s)} &  = &  \frac{10\pi\beta\alpha_e\alpha_g}{27s}(3-\beta^2),
\label{CrosSec}
\end{eqnarray}
where $\alpha_e = e^2 /(4\pi)$ and $\alpha_g = g^2 /(4\pi)$ are the fine structure constants for the electric and the magnetic charges, respectively, and  $\beta = \sqrt{1 - 4 M^2/s}$ is the monopole velocity.

\begin{figure}[!tbp]
  \centering
    \includegraphics[width=0.5\textwidth]{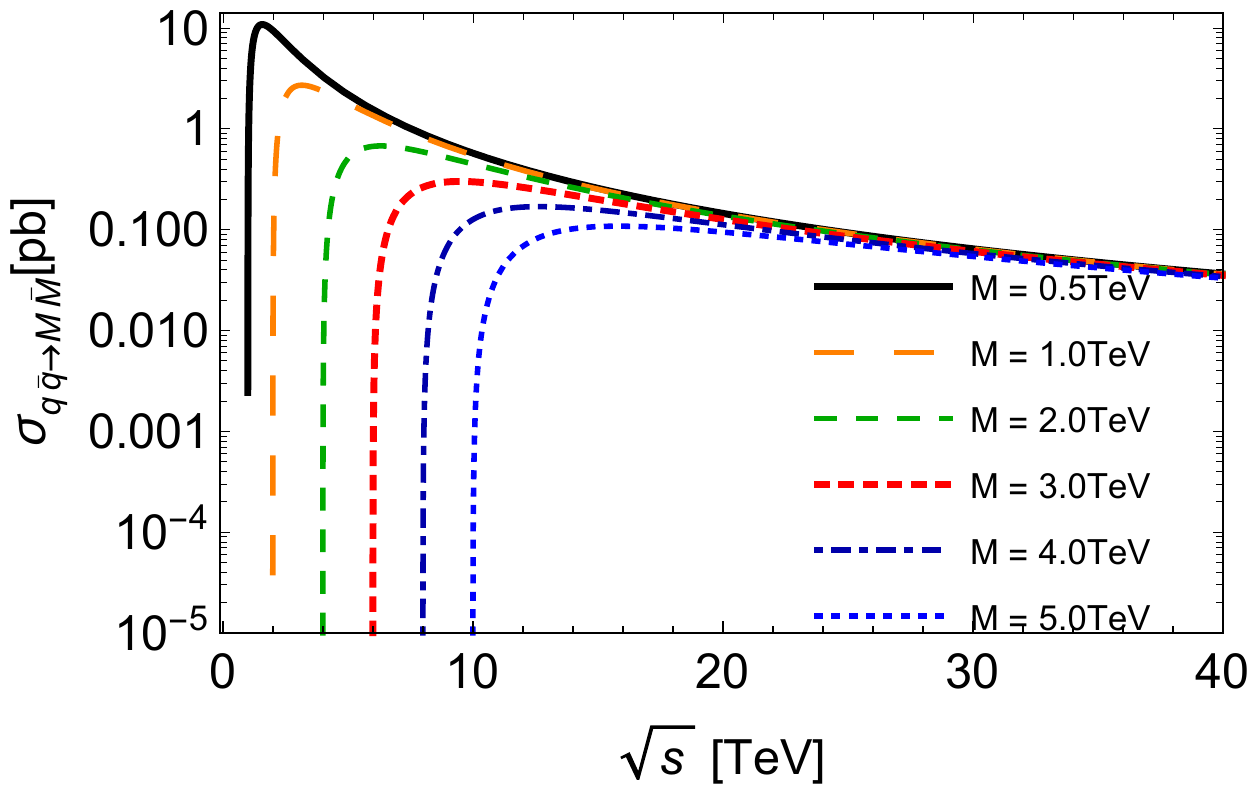} \\
    \includegraphics[width=0.5\textwidth]{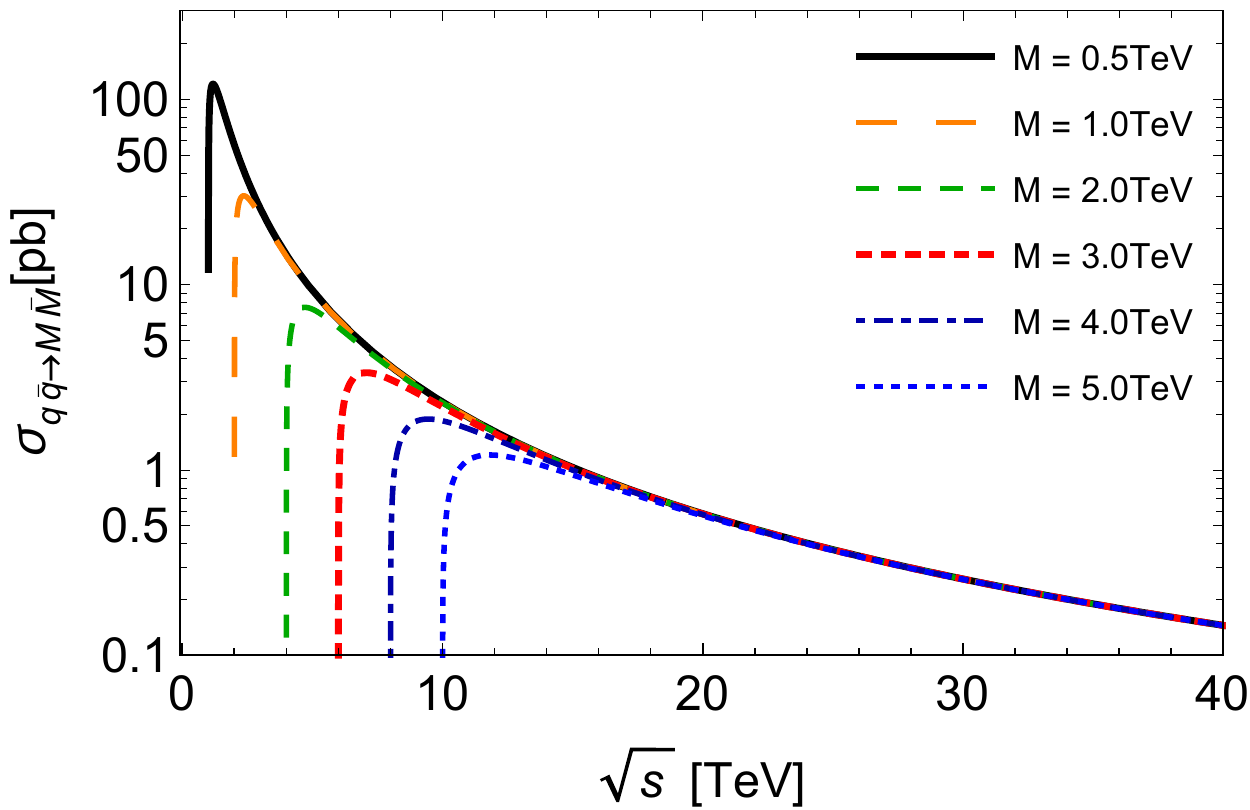}
    \caption{Total cross section for monopole -antimonopole pair production $q\bar{q}\rightarrow M\bar{M}$ as a function of the center-of-mass energy $\sqrt{s}$, with different values of the monopole mass $M$. Plots in the top and bottom panels describe spin-zero and spin-half monopoles}, respectively.
    \label{fig1}
\end{figure}

\begin{figure}[!tbp]
  \centering
    \includegraphics[width=1.0\linewidth]{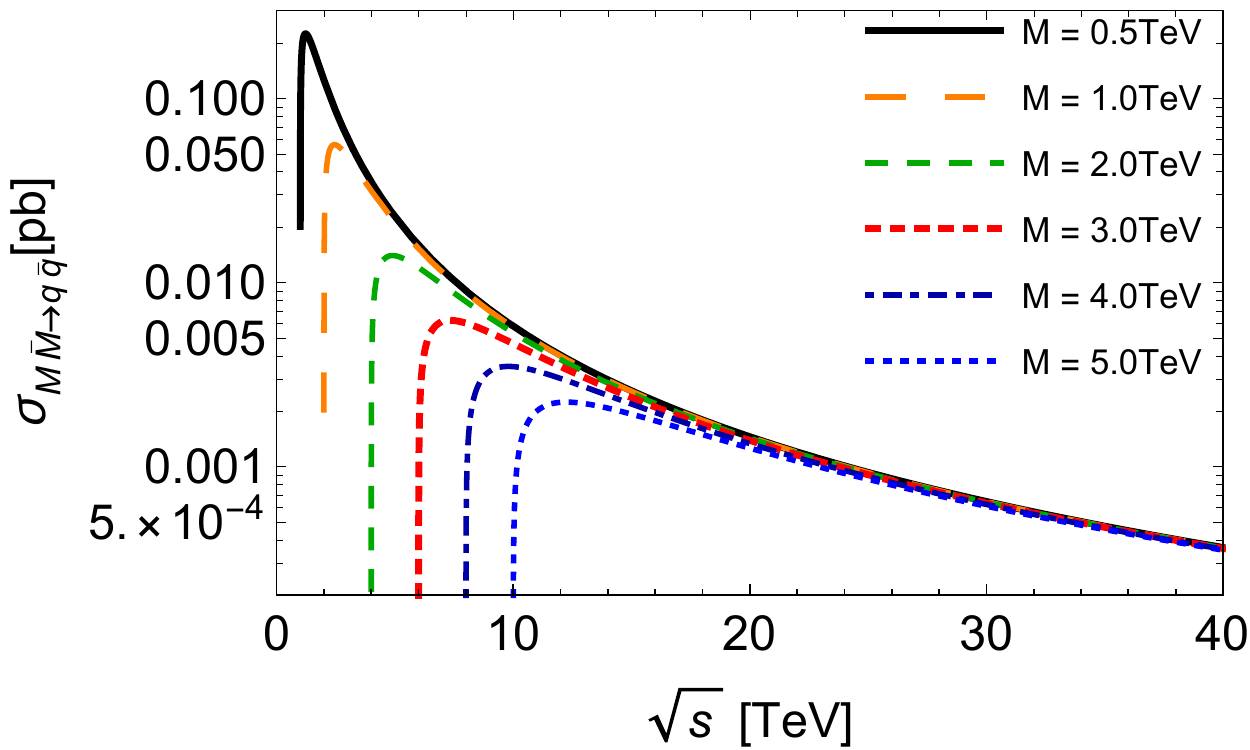} \\
    \includegraphics[width=1.0\linewidth]{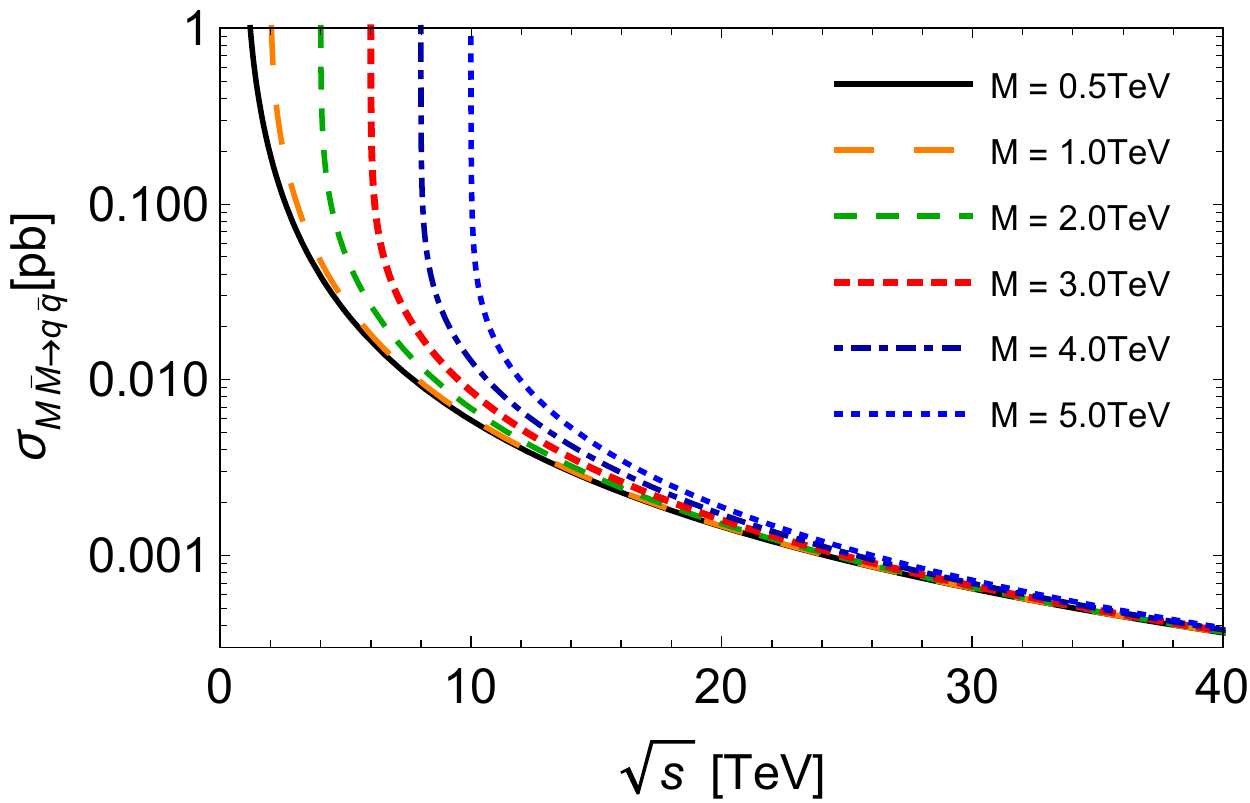}
    \caption{Total cross section for monopole -antimonopole} pair annihilation $M\bar{M}\rightarrow q\bar{q}$ as a function of the center-of-mass energy $\sqrt{s}$,  with different values of the monopole mass $M$. Plots in the top and bottom panelsdescribe spin-zero and spin-half monopoles, respectively.
    \label{fig2}
\end{figure}

The cross sections of the inverse processes, in which the monopole pair is annihilated via Drell-Yan processes, can be evaluated using the detailed balance relation, i.e.
\begin{eqnarray}
 g_{a}g_{b} |\vec{p}_{ab}|^2 \sigma_{ a b  \rightarrow c d } (s) = 
 g_{c} g_{d}  |\vec{p}_{cd}|^2   \sigma_{c d   \rightarrow a b } (s).
\label{detailedbalanceeq}
\end{eqnarray}
The total cross sections for the production and annihilation of monopole-antimonopole pairs are plotted in Figs. \ref{fig1} and \ref{fig2} as functions of the CM energy $\sqrt{s}$, for different values for the monopole mass between  0.5 and 5.0 TeV. The production cross sections are endothermic, and smaller values of $M$ engender higher cross sections. From Figs. \ref{fig1} and \ref{fig2}, we observe that the thresholds $\sqrt{s}_{min}$ increases with $M$, and that as the CM energy increases, the cross section $\sigma$ decreases for all processes, and that they trend towards the same behavior, almost indistinguishable at very high energies. At $\sqrt{s} \approx 15 $ TeV, for example, this effective approach suggests cross sections with magnitudes of the order $ \sim  10^{-1} $ pb and  $ 1 $ pb for spin-zero and spin-half, respectively. If we look now at the inverse processes plotted in Fig. \ref{fig2}: due to their exothermic nature, they might have a different behavior near the thresholds. This is the case for spin-half monopoles: the curves for $\sigma_{inv} ^{(1/2)}$ become very large at their respective thresholds, and quickly decrease with larger $\sqrt{s}$. For $\sigma_{inv}  ^{(0)}$, the use of Eq.~(\ref{CrosSec}) in (\ref{detailedbalanceeq}) yields a linear dependence with the factor $\beta$ and therefore a distinct behavior. More interestingly, from moderate CM energies onward, these inverse cross sections become smaller than those for production reactions.

\section{The thermally averaged cross section for production and absorption of monopoles\label{s3}}

Once the vacuum cross sections are obtained, the next step is to estimate the thermally averaged cross sections. Our interest in the present analysis  stems from the mechanisms of monopole production and absorption from the early universe onward. In the initial stages of its formation, the correspondence between the time evolution $t$ and the temperature $T$ was described approximately by $t$(s) $\approx$ $T^{-2}$(MeV) \cite{Zeldovich1981}, which means a larger temperature as we move earlier in time. Keeping in mind that the temperature of the system determines the collision energy, the relevant dynamical quantity is the cross section averaged over the thermal distribution. It might be interpreted as the convolution of the vacuum cross section with thermal momentum distributions of the colliding particles. The thermal average acts  suppresses part of the kinematical configurations close to the thresholds. In this sense, a strong threshold enhancement and the suppression observed in Figs. \ref{fig1} and \ref{fig2} might have no significance for our purposes. 

Thus, let us define the cross section averaged over the thermal distribution for a reaction involving an initial two-particle state going into two final particles $ab \to cd$ as~\cite{Koch}
\begin{eqnarray}
  \langle \sigma_{a b \rightarrow c d } \,  v_{a b}\rangle &  = & \frac{ \int 
    d^{3} \mathbf{p}_a  d^{3} \mathbf{p}_b \, f_a(\mathbf{p}_a) \,
    f_b(\mathbf{p}_b) \, 
    \sigma_{a b \rightarrow c d } \,\,v_{a b} }{ \int  d^{3} \mathbf{p}_a  d^{3}
    \mathbf{p}_b \, f_a(\mathbf{p}_a) \,  f_b(\mathbf{p}_b) }
\nonumber \\
& = & \frac{1}{4 x_a ^2 K_{2}(x_a) x_b ^2 K_{2}(x_b) }
\int _{z_0} ^{\infty } dz  K_{1}(z) \,\,
\nonumber \\
& & \times \sigma (s=z^2 T^2) \left[ z^2 - (x_a + x_b)^2 \right]
\nonumber \\
& &\left[ z^2 - (x_a - x_b)^2 \right],
\label{ThermalCrossSection}
\end{eqnarray}
where $v_{ab}$ represents the relative velocity of the two initial  interacting particles $a$ and $b$; $\sigma_{ab \to cd}$  denotes the cross sections   
evaluated formerly for the different reactions shown in
Figs.~\ref{fig1} and ~\ref{fig2}; the function $f_i(\mathbf{p}_i)$ is the Bose-Einstein   
distribution of particles of species $i$, which depends on the temperature    
$T$; $x _i = m_i / T$, $z_0 = max(x_a + x_b, x_c 
+ x_d)$; and  $K_1$ and $K_2$ the modified Bessel functions.

From Eq.~(\ref{ThermalCrossSection}), it can be presumed that for small values of $x_a, x_b$ the denominator reaches very high values; that is the case for finite temperatures and the relatively small masses of the quarks. Consequently, $ \langle \sigma_{a b \rightarrow c d } \,  v_{a b}\rangle$ acquires very small magnitudes for monopole production processes with respect to the suppression ones; within the accuracy used in our numerical calculations,   they become zero in most of the range of temperature considered. For this reason, hereafter we shall only consider the latter reactions.

To estimate the thermally averaged cross section for absorption processes $M\bar{M}\rightarrow q\bar{q}$, it is more convenient to write it as a function of 

\begin{eqnarray}
 x = M/T,\label{x-def}
\end{eqnarray} and the vacuum cross section in terms of the dimensionless variable $y = s / 4M^2$~\cite{Cannoni2016}, and after some algebra, one finds:
\begin{eqnarray}
  \langle \sigma_{M\bar{M}\rightarrow q\bar{q}} \, v_{M\bar{M}}\rangle ^{(S=0)} &  = & \frac{20\pi\alpha_g\alpha_e x}{27 M^2 K_{2}^2(x)} \left[ I_{-\frac{1}{2}} - I_{-\frac{3}{2}} \right],  \nonumber \\
  \langle \sigma_{M\bar{M}\rightarrow q\bar{q}} \,  v_{M\bar{M}}\rangle ^{(S=1/2)} &  = & \frac{10\pi\alpha_g\alpha_e x}{27 M^2 K_{2}^2(x)} \left[ 2 I_{-\frac{1}{2}} + I_{-\frac{3}{2}} \right], 
\label{ThermalCrossSection2}
\end{eqnarray}
where the function $I_{p}$ is given by
\begin{eqnarray}
  I_{p} &  = & \int _{1} ^{\infty }  dy \, (y-1) \, y^{p} \, K_{1}\left( 2 x \sqrt{y} \right). 
\label{Ip}
\end{eqnarray}
 This form therefore enables one to analyze the behavior of this relevant observable according to the ratio between the monopole mass and temperature of the medium. 
 
 \begin{figure}[!tbp]
  \centering
    \includegraphics[width=1.0\linewidth]{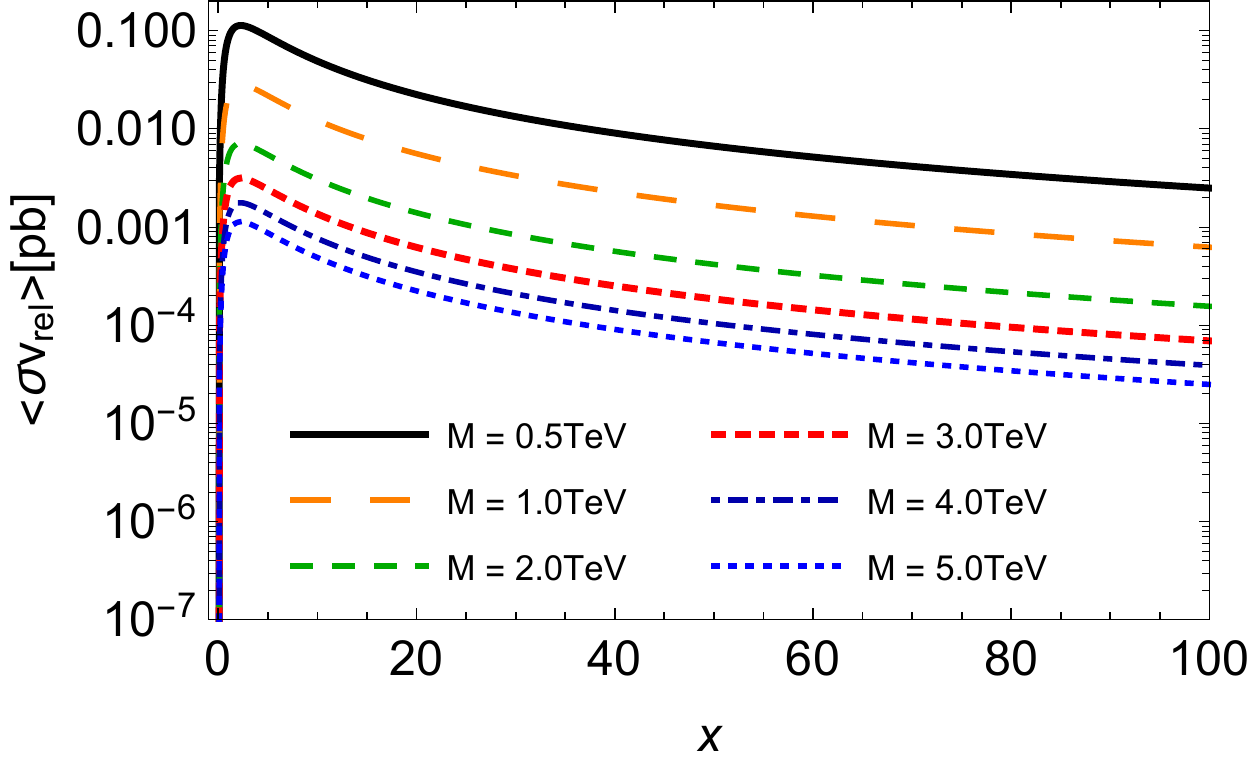} \\
    \includegraphics[width=1.0\linewidth]{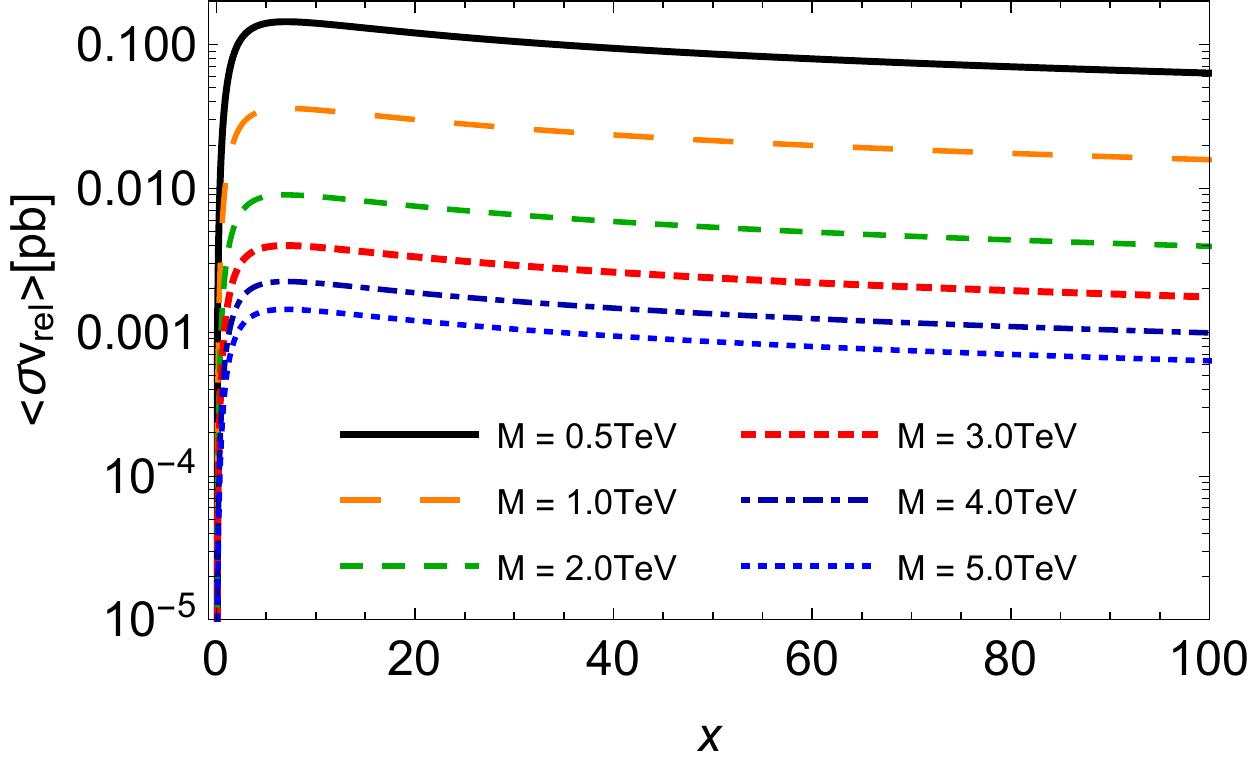}
    \caption{Thermally averaged cross sections for monopole pair annihilation $q\bar{q}\rightarrow M\bar{M}$ as a function of $x = M /T$, with different values of the monopole mass $M$. Plots in the top and bottom panels describe spinless and spin-half monopoles}, respectively.
    \label{fig3}
\end{figure}

In Fig.~\ref{fig3}, we plot the thermally averaged cross sections of Eq.~(\ref{ThermalCrossSection2}) as a function  of the parameter $x$ defined in Eq. \eqref{x-def}. Here again, higher magnitudes are achieved for reactions involving monopoles with smaller masses, and as the temperature decreases (i.e. the variable $x$ augments) this difference is kept nearly constant within the considered range of $x$. 
Also, these respective magnitudes are similar for both spin-zero and spin-half monopoles in the region near $x \sim 2$. We point out that, as we increase $x$, $ \langle \sigma_{M\bar{M}\rightarrow q\bar{q}} \, v_{M\bar{M}}\rangle $  for spin-zero drops faster than that for spin-half; indeed, the latter suffers just a very slight decrease in the studied range. In other words, the decreasing of temperature engenders a rate of annihilation for scalar monopoles smaller than the one for fermionic monopoles. In the context of our effective approach, it might be interpreted as a theoretical evidence of a more pronounced stability for spin- zero and for heavier monopoles.  
It is noteworthy that the difference between the magnitudes of thermally averaged cross sections for the annihilation reactions and the production reactions might, in principle, play an important role in the search for monopoles in future heavy-ion colliders and  in the evolution of the monopole abundance of cosmic origin. We will investigate the evolution of this abundance in the next section.

\section{Relic Abundance\label{s4}}

\subsection{Framework}

In this section, we estimate the evolution of the relic monopole abundance during the expansion of the universe.  In general, one can use the thermally averaged cross sections estimated in the previous section as inputs in the momentum-integrated evolution equation abundance to assess the gain and loss terms, due to  production and annihilation processes, respectively. But additionally, we will make use of the so-called ``freeze-out theory''~\cite{Cannoni2015,Cannoni2016,Weinberg1977,Gondolo1991}. This framework has been employed in several scenarios, like in the analysis of dark matter. Its basis relies on the assumption that the early universe expanded and cooled down, reaching a specific temperature, the ``freeze-out temperature'', at which the production and annihilation rates of the stable particles become uniform. Past this stationary point, we we are left with a residual number of particles, i.e. the relic abundance. The instant of the freeze-out for a given particle depends on its mass and interactions; we designate it by $x=\Tilde{x}$. From this perspective, the interactions of massive and stable monopoles with other particles might explain the monopole phenomenology.

We begin with the rate equation governing the evolution of the relic number density $n$~\cite{Cannoni2015,Weinberg1977}:
\begin{equation}
    \frac{dn}{dt}+\frac{3\dot{R}}{R}n=-\left<\sigma_{M\bar{M}\rightarrow q\bar{q}}v_{M\bar{M}} \right>(n^2 - n_0^2) ,
    \label{rateequation}
\end{equation}
where $\left<\sigma_{ann} v \right>$ is the thermally averaged cross section for the annihilation process (i.e. $M\bar{M}\rightarrow q\bar{q}$ in our case); $n_0$ is the number density in the thermal equilibrium; $R$ is the cosmic scale factor related to the Hubble parameter $R$ through $H =\dot{R}/R = \sqrt{8\pi G \rho/3}$, with $G=1/M^2_P$ being the cosmological constant  ($M_p = 1.22\times10^{16}$ TeV is the Planck mass) and $\rho=\pi^2 g_{\rho}T^4  / 30$ the total energy density of the universe  ($g_{\rho}$ counts the relativistic degrees of freedom contributing to the energy density).

Remembering that $\left<\sigma_{M\bar{M}\rightarrow q\bar{q}}v_{M\bar{M}}\right>$  in Eq.~(\ref{ThermalCrossSection2}) has been given as a function of the parameter $x$ defined in Eq. \eqref{x-def}, it is convenient to write Eq.~(\ref{rateequation}) in a different form. Given the entropy density of the universe  $s = (2\pi^2/45)g_sT^3$, where $g_s$ refers to relativistic degrees of freedom associated to the the total entropy density~\cite{Cannoni2015,Cannoni2016,Weinberg1977,Gondolo1991}, then the hypothesis that the expansion proceeds adiabatically imposes the conservation of the total entropy per comoving volume $S=R^3s$. In consequence, the quantity $Y \equiv N/ S = n/s$ appears as an appropriate observable to be employed in the present analysis. We obtain it by dividing Eq.~(\ref{rateequation}) by $S$. Thereafter, the variable is changed by employing the correspondence $d/dt\rightarrow Hxd/dx$, enabling one to rewrite the rate equation for the relic abundance as function of $x$ in the form, 
\begin{equation}
    \frac{dY}{dx}=\frac{C}{x^2}\left<\sigma_{M\bar{M}\rightarrow q\bar{q}}v_{M\bar{M}}\right>(Y^{2}_{0}-Y^2), 
    \label{rateequation2}
\end{equation}
where $Y_0 = 45/(4\pi^4)(g/g_s)x^{2}K_{2}(x)$ is the initial equilibrium abundance ($g = 1$ or 2, for scalar or fermionic monopoles, respectively), and  $C=\sqrt{\frac{\pi}{45}}M_{P}M\sqrt{g_{*}}$, with 
\begin{equation}
\sqrt{g_{*}}= \frac{g_s}{\sqrt{g_{\rho}}} \left( 1+ \frac{T}{3}\frac{d(\ln g_s)}{dT} \right)
    \label{gstar}
\end{equation}
depending on the relativistic degrees of freedom $g_{\rho}$ and $g_s$.  As pointed out for example in Ref.~\cite{Steigman:2012nb}, the degrees of freedom remain almost constant over a limited range of temperature. Consequently, it sounds reasonable to neglect their temperature dependence. Furthermore, $g_{\rho}$, and $g_s$ differ only when there are relativistic particles present that are not in thermodynamic equilibrium with the photons, which is not our situation here. Thus, we take $g_s = g_{\rho} =  100$, and $\sqrt{g_{*}} = \sqrt{g_{s}}$~\cite{Cannoni2016}.

At this point, let us briefly elucidate a crucial  assumption that we utilized: as done in the literature (for example, Refs. \cite{Cannoni2015,Cannoni2016}), we conjecture that for a given monopole mass, our effective approach predicts the total annihilation cross section averaged over the thermal distribution, and therefore, it is possible to estimate the actual monopole abundance at a given stationary point $\Tilde{x}$ , which is then interpreted as the true initial condition for the rate equation in Eq.~(\ref{rateequation2}). This enables one to get the asymptotic value for the abundance at large $x$. In other words, the true abundance becomes dependent on the point $\Tilde{x}$. Thereby, the relic abundance $Y(x)$ can be defined by piecewise function 
\begin{eqnarray}
 Y(x)=\begin{cases} 
      Y_1(x) , \,  & x\leq \Tilde{x}, \\
      Y_2(x) , \,  & x\geq \Tilde{x},
   \end{cases}
    \label{piecewise}
\end{eqnarray}
where the function $Y_1(x)$ describes the evolution before $\Tilde{x}$ and is given by 
\begin{equation}
    Y_1(x) = \sqrt{Y^2_0 - \frac{x^2}{C\left<\sigma_{M\bar{M}\rightarrow q\bar{q}}v_{M\bar{M}}\right>}\frac{dY_0}{dx}}, 
\label{Y1}
\end{equation}
while  $Y_2(x)$ is the actual abundance governed by the rate equation~(\ref{rateequation2}) for $ x \geq \Tilde{x}$, with the true initial condition $Y_2(\Tilde{x})  =  Y_1(\Tilde{x})$. Hence, the differential evolution in temperature starts at $x = \Tilde{x}$, where $Y(x)$ is continuous and differentiable. 

The final information needed is the stationary point $ \Tilde{x}$.  We determine it by considering the quantity $\Delta = Y - Y_0 $, the amount of abundance distant from equilibrium. Then, the matching point between the solutions in the two regions of $x$ can be estimated by means of $\Delta (\Tilde{x}) = c Y_0  (\Tilde{x})$, where $c$ is a numerical factor determined by the numerical solution of the rate equation. Different values for $c$ have been proposed in  the literature in distinct physical scenarios (see Ref.~\cite{Cannoni2015} for a detailed discussion). As a first attempt, here we adopt the condition $\Delta (\Tilde{x}) = Y_0  (\Tilde{x})$,  yielding  $ Y_2 (\Tilde{x}) = Y_1 (\Tilde{x}) = 2 Y_0 (\Tilde{x})$. 
We remark that a similar choice is employed in Refs.~\cite{Zeldovich,Weinberg1977} for the approximate initial condition at the freeze-out point in other contexts. 
Hereupon, rewriting $Y_1(x)$ in Eq.~(\ref{Y1}) in the form $Y_1(x) \equiv (1 + \delta (x) ) Y_0(x)$, where 
\begin{equation}
\delta (x)  =  \sqrt{1 - \frac{x^2}{C\left<\sigma_{M\bar{M}\rightarrow q\bar{q}}v_{M\bar{M}}\right> Y_0^2}\frac{dY_0}{dx}} - 1 , 
\label{delta}
\end{equation}
we obtain, for the assumed condition above, $ \delta (\Tilde{x}) = 1$, which is equivalent to 
\begin{equation}
- \left. \frac{1}{Y_0}\frac{dY_0}{dx} \right|_{ x = \Tilde{x}} =   \left[  3\frac{C}{x^2}\left<\sigma_{M\bar{M}\rightarrow q\bar{q}}v_{M\bar{M}}\right> Y_0  \right]_{ x = \Tilde{x}}. 
\label{cond}
\end{equation}

Finally, it is also useful to analyze the chemical potential of monopoles. It is related to the  ``affinity'', defined as $\mathcal{A} = - \sum _i \nu_i \mu_i$, where $\mu_i$ are the chemical potentials of the species of type $i$, and $\nu_i$ are the stoichiometric coefficients, which are assumed to be negative for monopoles and antimonopoles, yielding $\mathcal{A} = 2 \mu $~\cite{Cannoni2015}. If we assume that it is related to the $log$ of the ratio between the production and annihilation rates, i.e. the two terms in the right-hand side of the rate equation in~(\ref{rateequation}), then one can write the chemical potential as
\begin{equation}
    \mu=\frac{M}{x}\ln{\left( \frac{Y}{Y_0} \right)}.
\label{chempot}
\end{equation}

\subsection{Numerical results and discussion}

We can now estimate the relevant quantities introduced previously. Firstly, we find the magnitudes of $\Tilde{x}$ required by the condition in Eq. ~(\ref{cond}). They are shown in Table~\ref{tab1} for different values of the monopole mass. We notice  that $\Tilde{x}$ decreases as $M$ increases, due to the $M$-dependence of the thermally averaged cross section, $C$ and $Y_0$, encoded in Eq.~(\ref{cond}). This means that heavier monopoles reach the equilibrium at higher temperatures; that is, in older epochs. Let us also remark that, although in principle the dependence on the spin is present in the cross section and degenerescence factor $g$ in the condition (\ref{cond}), the numerical solutions for $\Tilde{x}$ do not present sizable  differences between spin-zero and spin-half, suggesting that they would have left a relic abundance at the same moment of evolution of the universe, viz. the  same temperature $\Tilde{T}$ engendered by the stationary point $\Tilde{x}$.
Given the rather small values of $M$ we are treating, in the last array of Table~\ref{tab1} we also included a much higher mass, chosen accordingly to Refs.~\cite{Mavromatos:2020gwk,Arunasalam:2017eyu} where that higher mass is related to the upper limit imposed by the nucleosynthesis constraints on the abundance of relic monopoles.

\begin{center}
\captionof{table}{Magnitudes of temperature $\Tilde{T}$ and $\Tilde{x} = m / \Tilde{T} $ obtained from the condition in Eq.~(\ref{cond}) for different values of the monopole mass.  \label{tab1}}
\begin{tabular}{c|c|c}
\hline                         
\hline                         
\vspace{2pt}
Mass (TeV) & $\Tilde{T}$ (TeV) & $\Tilde{x}$ 
\\ 
\hline                         
0.5        & 0.02       & 28.76     \\
1.0        & 0.04       & 28.09     \\
2.0        & 0.07       & 27.42     \\
3.0        & 0.11       & 27.03     \\
4.0        & 0.15       & 26.76     \\
5.0        & 0.19       & 26.54     \\
$10^4$       & 519.49     & 19.25     \\
\hline                         
\hline                         
\end{tabular}
\end{center}

\begin{figure}[!tbp]
  \centering
    \includegraphics[width=1.0\linewidth]{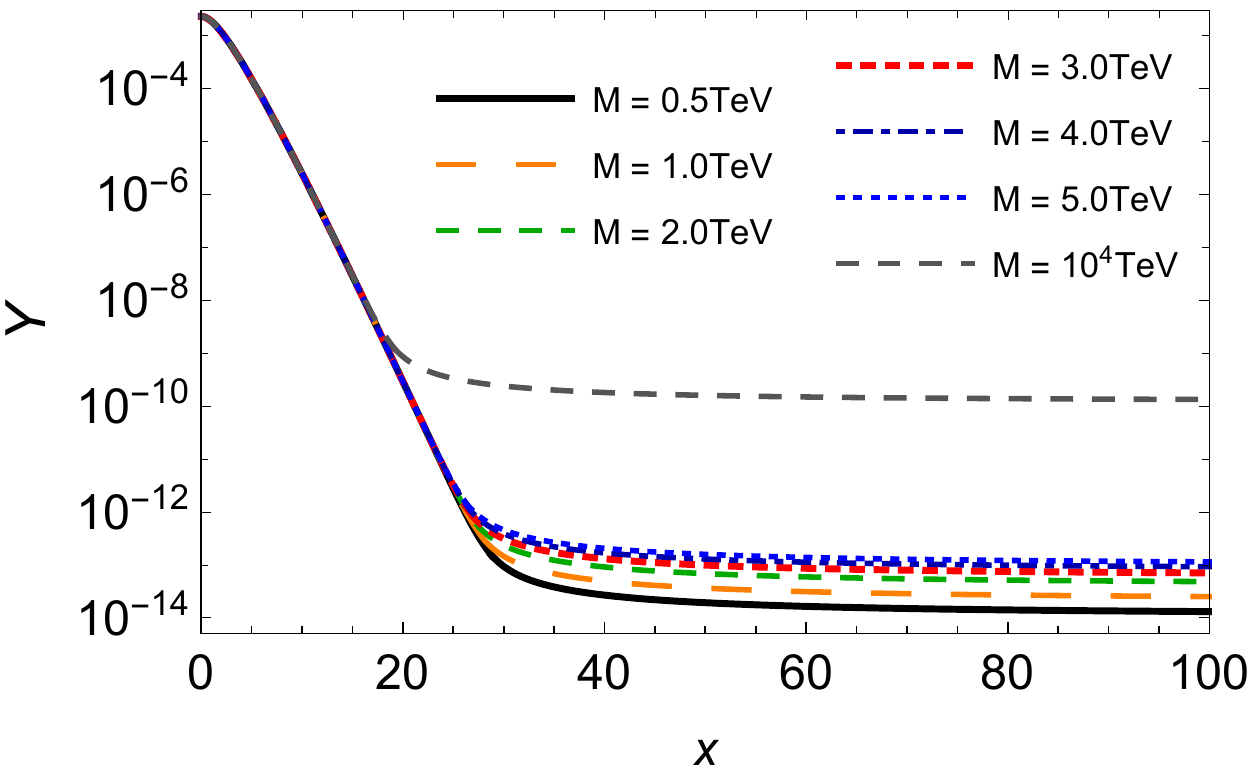} \\
    \includegraphics[width=1.0\linewidth]{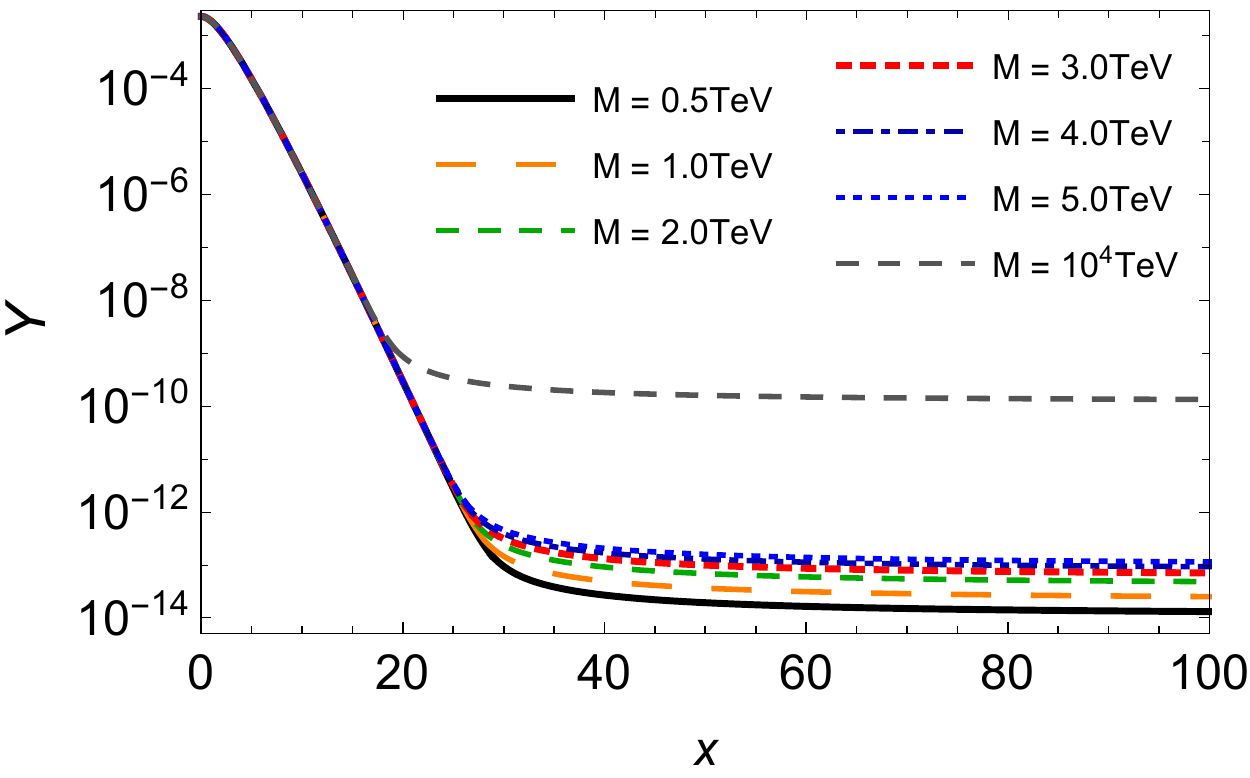}
    \caption{The relic abundance $Y(x)$ for monopole as a function of $x = M /T$, according to Eq.~(\ref{piecewise}), taking different values of the monopole mass $M$. Plots in the top and bottom panels: cases of spin-zero and spin-half monopoles, respectively.}
    \label{fig4}
\end{figure}

 Fig.~(\ref{fig4}) shows plots of the relic abundance $Y(x)$ for spinless and spin-half monopoles as a function of $x = M /T$, according to  the piecewise function Eq.~(\ref{piecewise}), with different values of the monopole mass $M$. We notice that before the stationary point, the abundance evolves almost in the same way for different values of the monopole mass. From larger values of $\Tilde{x}$, the curves become perceptibly distinct. Because  heavier monopoles experience smaller stationary points, they achieve the equilibrium at earlier stages of the expansion (i.e. higher temperatures). No less important is that the increase of $M$ produces higher values of the relic abundance. Specifically, an augmentation of one order of magnitude in $M$ yields roughly a similar growth in the magnitude of $Y$ at large $x$. Besides, the results suggest that the abundance does not behave differently for spin-zero and spin-half relic monopoles. 

From the quantitative point of view, the values estimated for the relic abundance are clearly high, in view of the effective formalism and the magnitudes of the couplings and quantities taken. In that regard, the density of relic monopoles is naturally modified for different set of values for the relevant parameters, e.g.  $g_D, \sqrt{g_{*}}, \sqrt{g_{s}}$ and so on. Notwithstanding, since the purpose here is not to make very accurate predictions, we postpone a more detailed analysis of this issue for future studies.

It is useful to compare these findings to others in the literature. For instance, in the seminal paper~\cite{Turner1982}, the thermal production of magnetic monopoles has been calculated by using the monopole-antimonopole annihilation cross section and  the  detailed balance, and starting with an initial vanishing abundance. It has been found that the final abundance of monopoles depends upon $x_i^3 e^{-2 x_i} $, with $x_i = M / T_t$ ($ M \sim 10^{13}$ TeV and $T_i$ being the initial temperature of the production). Obviously, it is hard to perform a quantitative comparison because of the different purposes and values of the quantities considered (e.g. the mass used is very large with respect to the range considered here). But we remark some qualitative features of the approach employed in Ref.~\cite{Turner1982}: the thermally averaged cross section has a general dependence on the temperature as $T^{-2}$, 
and the relic abundance decreases monotonically for large $x$. These outcomes are clearly in contrast with those reported above. The main reason comes from the fact that the author of Ref. ~\cite{Turner1982} uses only the solution of rate equation with a negligible initial abundance in the whole range of $x$, in some sense considering only the behavior of the function $Y_1$, in opposition to the piecewise function in Eq.~(\ref{piecewise}).

\begin{figure}[!tbp]
  \centering
    \includegraphics[width=1.0\linewidth]{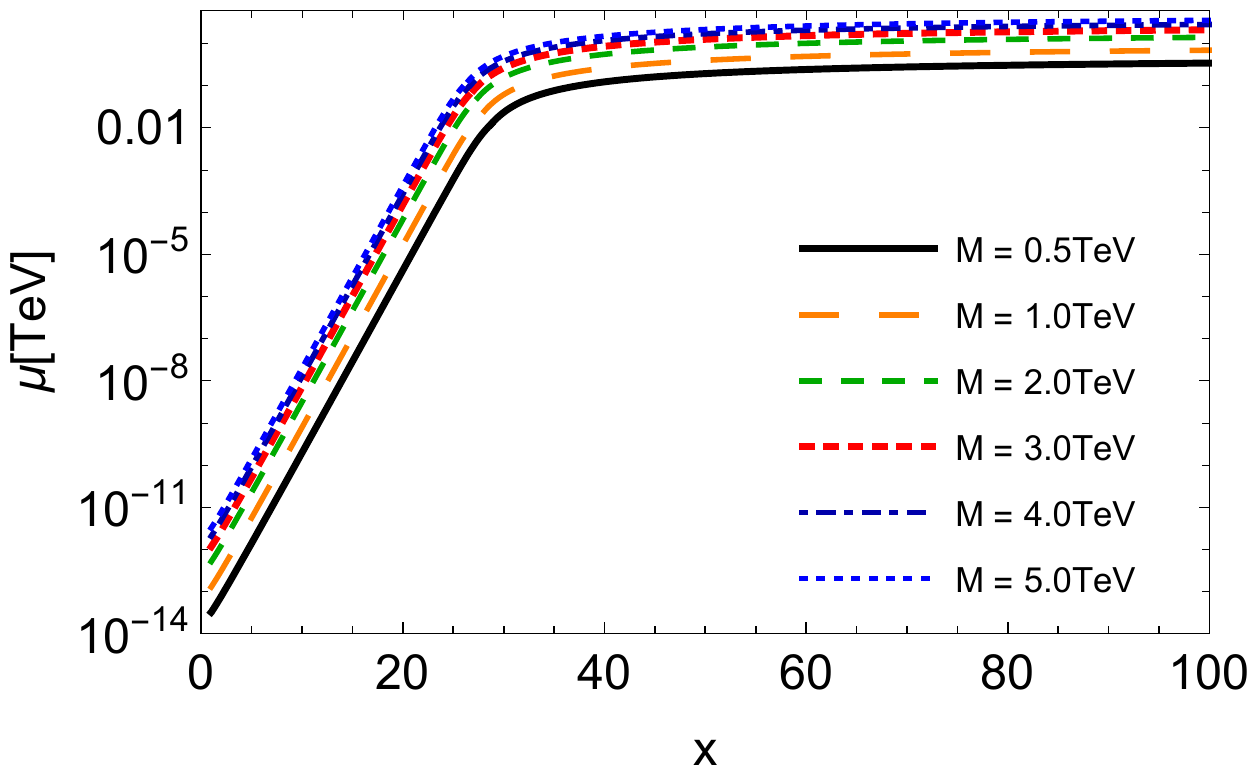} \\
    \includegraphics[width=1.0\linewidth]{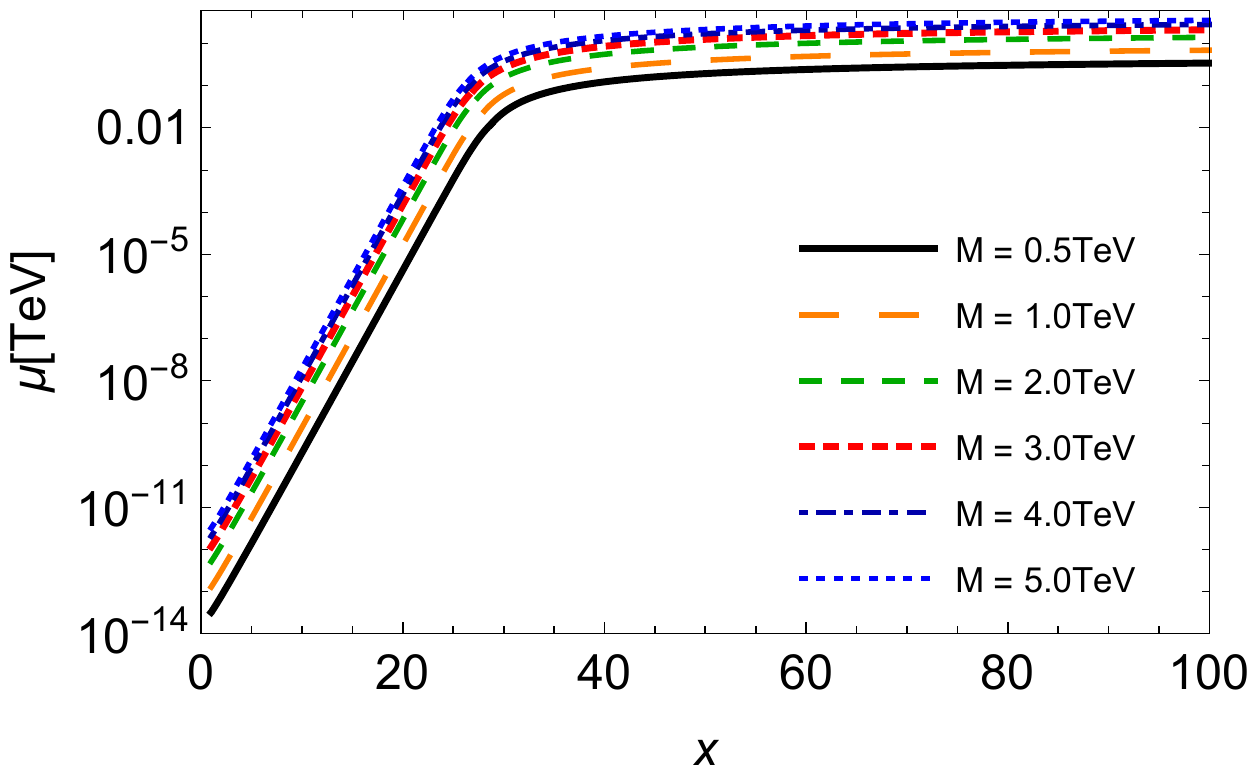}
    \caption{The chemical potential for monopoles as a function of $x = M /T$, according to Eqs.~(\ref{chempot}) and~(\ref{piecewise}), taking different values of the monopole mass $M$. Plots in the top and bottom panels: cases of spin 0 and $1/2$, respectively.}
    \label{fig5}
\end{figure}

We finish this analysis by showing in Fig. (\ref{fig5}) the plots of the evolution of the chemical potential as a function of $x = M /T$ for spinless and spin-half monopoles, according to Eqs.~(\ref{chempot}) and~(\ref{piecewise}), for different values of $M$. Since a non-zero chemical potential means a process out of chemical equilibrium, we see that at earlier stages of the universe, $\mu $ increases by several orders of magnitude, but as $x$ increases and the rate in  the definition of $\mu $ decreases, this yields a moderate growth. At larger $x$, i.e. smaller temperatures, this rate is very small, and $\mu$ saturates to an asymptotic value and the abundance tends to be constant. Interestingly, the asymptotic value is just the monopole mass, similarly to the findings in Ref. ~\cite{Cannoni2015}.

\section{Concluding Remarks\label{s5}}

In this work, we have revisited the thermal production and annihilation of monopoles and their relic abundance, exploiting the monopole phenomenology described by the effective field theory recently proposed in the description of monopole pair production via Drell-Yan and photon fusion processes \cite{Baines2018}. To this end, we have used the vacuum cross sections for the Drell-Yan reactions derived within the mentioned framework to estimate the cross section averaged over the thermal distribution associated to other particles that constitute the hot medium where the monopoles propagate. In the considered range of monopole mass with spin zero and spin half, our results suggest that the thermally averaged cross sections for the pair production are highly suppressed, while at higher temperatures, these cross sections for the annihilation of lighter pairs  reach larger values. Besides,  for smaller temperatures, the rate of annihilation for scalar monopoles is smaller than for fermionic monopoles, which might be interpreted as a theoretical evidence of a more pronounced stability for spin- zero and heavier monopoles. This might be relevant in the search for monopoles in future heavy-ion colliders and of cosmic origin.

With our thermally averaged cross sections as inputs, we have also employed the effective approach of Ref. \cite{Baines2018} to  describe the evolution of the monopole abundance by extending a framework utilized previously for the number density of dark matter candidates ~\cite{Jungman,Zeldovich,Weinberg1977,Cannoni2015}. 
From our results, we can infer that heavier monopoles experience smaller stationary points, achieving the equilibrium at earlier stages of the expansion, i.e. at higher temperatures. In  addition, monopoles with larger mass produce higher values of the relic abundance. Specifically, an increase by one order of magnitude in $M$ yields roughly a similar growth in the magnitude of $Y$ at large $x$. Our results also suggest that the abundance does not behave differently for spinless and spin-half relic monopoles. 

To conclude, we emphasize that our general objective was to study the implications of the effective field theory  of Ref. \cite{Baines2018} on some aspects of the monopole phenomenology. Clearly, more accurate analyses are needed to improve the present investigation such as, for example, the impact of our distinct assumptions on the values of the relevant parameters, the inclusion of other variables of interest (e.g. magnetic background), and so on. We postpone them for future discussions. 
	 
\begin{acknowledgements}

L.M. Abreu is grateful to the Brazilian funding agencies CNPq (309950/2020-1 and 400546/2016-7) and FAPESB (contract INT0007/2016).  M de Montigny thanks the Natural Sciences and Engineering Research Council of Canada (NSERC) for financial support (grant number RGPIN-2016-04309).  P.P.A. Ouimet and M de Montigny are also grateful for funding from NSERC (grant number sAPPJ-2019-00040).

\end{acknowledgements}


\end{document}